\documentstyle[preprint,aps,12pt]{revtex}
\tightenlines
\begin{document}

\draft

\title{Sensitivity to the pion-nucleon coupling constant \\
         in partial-wave analyses of \\
         $\pi N\to \pi N$, $NN \to NN$, and $\gamma N\to \pi N$ }
\author{R.A. Arndt\footnote{e-mail: arndtra@said.phys.vt.edu} }
\address{Department of Physics, Virginia Polytechnic Institute and State
University, Blacksburg, VA 24061}
\author{I.I. Strakovsky and R.L Workman}
\address{Department of Physics, The George Washington University, Washington,
DC 20052}
\author{M.M. Pavan}
\address{Massachusetts Institute of Technology, 77 Massachusetts Ave.
Cambridge, MA 02139}

\date{\today}
\maketitle

\begin{abstract}

We summarize results obtained in our studies of the pion-nucleon coupling 
constant. Several different techniques have been applied to
$\pi N$ and $NN$ elastic-scattering data, and the existing database for
single-pion photoproduction. The most reliable determination 
comes from $\pi N$ elastic scattering. The sensitivity in this reaction was 
found to be greater, by at least a factor of 3, when compared with analyses 
of $NN$ elastic scattering or single-pion photoproduction.

\end{abstract}
\vspace*{0.5in}

\narrowtext
\centerline{\bf 1. Introduction}

A number of groups have recently extracted the pion-nucleon coupling
constant ($g^2 / 4\pi$), mainly from analyses of $\pi N$ and $NN$ 
elastic scattering. In the $\pi N$ case, results have been obtained
via dispersion relations\cite{vpi_pin}, through the fitted couplings
of a tree-level model\cite{matsinos}, and from the long-range part
of a potential approach\cite{timmermans}. Values of $g^2 / 4\pi$ 
have been determined from the one-pion-exchange part of the $NN$ 
interaction in fits\cite{vpi_nn,nijmegen} 
to either the ``full'' low-energy database or restricted kinematic
regions expected\cite{teo} to be more sensitive to this coupling.

In order to determine which data are most sensitive
to $g^2 / 4\pi$, we map $\chi^2$ versus $g^2 /4 \pi$
for a variety of energy-dependent partial-wave analyses. The parabolic 
``widths'', near the resulting minima, 
are taken as a measure of the ``uncertainty'' 
in the determinations. In Section 2, we give results from our analysis of 
$\pi N$ elastic scattering to $T_{\pi}$=2.1 GeV. In Section 3, we discuss
a number of global fits of $NN$ elastic scattering data to $T_{lab}$=400 MeV.
In Section~4, we consider our current analysis of 
pion-photoproduction taken to $E_{\gamma}$=500 MeV, as well as mappings
extended to $E_{\gamma}$=2 GeV. This reaction also shows a sensitivity
to the value of $g^2 /4\pi$  Finally, in Section~5, we give a summary of our
conclusions.

\vskip .3cm
\centerline{\bf 2. $\pi N$ elastic analysis to 2.1 GeV}

In our most recent analysis of this reaction, we have added fixed-t
dispersion-relation constraints on the C$^+$ amplitude and forward
dispersion-relation constraints on the E$^{\pm}$ amplitudes in order to
facilitate an extraction of the $\sigma$-term. A mapping of 
$\chi^2$, as a function of $g^2 / 4\pi$,
reveals a very deep minimum at 13.73(0.01). This is, of course,
the charged $\pi NN$ coupling, and we shall refer to the (0.01) as a mapping
error. 

The coupling constant must be extracted through the 
evaluation of a dispersion relation and this leads to an extraction
error. The Hamilton dispersion relation for the B$^+$ amplitude was
used by Carter, Bugg and Carter\cite{Bugg} to extract the long-standing value
of $g^2/4 \pi$=14.28. This is the value used in the 
Karlsruhe-Helsinki\cite{ka84}
and Carnegie-Mellon$-$Berkeley analyses\cite{CMB} of the late 1970's. 
If we use the Karlsruhe solution, KA84, in the Hamilton dispersion
relation, for energies between 100 and 600 MeV (Tlab), we obtain a 
value of $g^2/4 \pi$=14.40(0.23), where (0.23) is the RMS deviation from
a mean value of 14.40. Evaluating this dispersion relation for
our current solution, SP99, results in a value of $g^2/4 \pi$=13.73(0.03).
The (0.03) we shall refer to as an extraction error and, it is 
evident that this is the primary source of uncertainty.
A more complete description of our solution, SP99, is given by
Marcello Pavan in a separate contribution to these proceedings.

\vskip .3cm
\centerline{\bf 3. $NN$ elastic analysis}

Sensitivity to the coupling constant is built into our representation
for NN elastic scattering in the following fashion.
The scattering amplitude (suppressing spin) is given by:

\begin{equation}
  f=f_{\rm coul} + B_p + \sum_{l=0}^8 (2l+1)P_l(Z)
  f(l,E) {\rm e}^{2i\phi_l} ,
\end{equation}
where $B_p$ gives the peripheral Born amplitude (the Born term minus the 
partial waves between $l=0$ and $l=8$), and
\begin{equation}
  f(l,E)=K(l,E)/(1-iK(l,E)) ,
\end{equation}
where
\begin{equation}
  K(l,E)= {\rm Born}_l + \sum_{i} \alpha_i A_{li} .
\end{equation}

The Born term contains a t-channel pole proportional to $g_0^2/4 \pi$ 
and a u-channel pole proportional to $g_c^2/4 \pi$.
The expansion bases, $A_{li}$, are chosen to have a left-hand cut that starts
at the $2\pi$ threshold and have the correct 
threshold behavior\cite{vpi_nn}.

The sensitivity of the fitting scheme to $g^2/ 4\pi$ clearly 
depends upon the amount of phenomenology
(number of $\alpha_i$ terms used). In order to estimate the effect of
phenomenological contributions, we considered two different solutions
fitted to the current GW data base below 400 MeV. (The upper limit
was chosen to keep the analyses essentially elastic.) 
Solution SP40 was obtained by fitting 57
parameters for waves with $J < 7$, while SG40 utilized 48 parameters for
waves with $J < 5$. In all of our mappings, we assumed that there was
no charge splitting; $g_0=g_c$. 

Tables I and II summarize our results from the $\chi^2$-mapping 
of these solutions. Uppsala data at 96 and 162 MeV have been used in
conjunction with the GW data base. The columns labeled PPx and
NPx are mappings which exclude the Uppsala cross sections. Solution
SG40 shows a greater sensitivity, as expected, with a consequent
increase in $\chi^2$ of around 350 for the 9 eliminated search parameters.
This might raise questions concerning the actual value of $g^2/4 \pi$
extracted in such analyses. However, our focus is on sensitivity which
increases by about 50\% as we go from SP40 to SG40. We make no attempt
to address the issue of possible charge-splitting, but it should be
pointed out that $pp$ data depend only upon $g_0^2/4 \pi$ whereas $np$ data
depend upon both coupling constants.

The focus of the workshop has been on $np$ charge-exchange cross-sections,
where there is an obvious and strong dependence upon the charged
coupling constant, but it is important to realize which other data 
are sensitive
to the coupling constant. In order to partially resolve that issue, we
used solution SG40 to map a data base which excluded
ALL $np$ cross sections on the assumption that total cross sections
would set the scale of the interaction, while the angular shape
would be constrained by the abundant
spin data in the $np$ data base. The resultant solution, renamed SX40,
was used to generate the results in Table III.

While this is certainly an inappropriate 
determination of the coupling constant, 
it does reveal a substantial sensitivity in the spin data, and suggests
that their inclusion, in any extraction of $g^2/ 4\pi$ from $np$ elastic
scattering data, will alter the final result. 

\vskip .3cm
\centerline{\bf 4. Analysis of pion-photoproduction data}

Sensitivity to $g^2/ 4\pi$ is built into the representation used to
parameterize pion photoproduction in manner similar to 
$NN$ elastic scattering. The scattering amplitude $M$ is basically
taken to be
\begin{equation}
M = B_p + \sum_{l=0}^{l_{\rm max} } m_l a_l(\theta ) ,
\end{equation}
where
\begin{equation}
B_p = Born - \sum_{l=0}^{l_{\rm max} } b_l a_l(\theta ) .
\end{equation}
The contribution, $B_p$, is the residual Born contribution not included
in the fitted multipole ($m_l$) amplitudes, 
$b_l$ is the projected Born term, and $a_l(\theta )$ represents
the expansion basis.
The amplitudes, $m_l$, are then expressed as
\begin{equation}
m_l = \left( b_l + c_l \right) \left( 1 + i T_{\pi N} \right)
 + d_l T_{\pi N} ,
\end{equation}
$c_l$ and $d_l$ being structure functions of energy which are fitted
to the available data and $T_{\pi N}$ being the elastic $\pi N$ scattering
amplitude. This form ensures that Watson's theorem is satisfied  below
the 2-pion production threshold. Dependence upon the coupling constant
is contained in $T_{\pi N}$, as described in Section~2, and in the Born
terms, $B_p$ and $b_l$. 

Table~4 characterizes a fit to pion-photoproduction data from threshold
($E_{\gamma }$ $\approx$ 145 MeV) to 500 MeV. It reveals a sensitivity
about 50\% weaker than that obtained from the $np$ mapping of solution
SG40 (an error of $\pm 0.13$ versus $\pm 0.09$). 
A more extensive study of the full data base, which extends from
threshold to about 2 GeV, is summarized in Fig.~1, where the maximum
value of $E_{\gamma }$ was varied from 500 MeV to 2 GeV. An average
of the extracted $g^2 / 4\pi$ values reveals a sensitivity of about
($\pm 0.14$). 

\vskip .3cm
\centerline{\bf 5. Conclusions}

We have looked at three fundamental
scattering reactions having strong signatures for the
pion-nucleon coupling constant, $g^2 / 4\pi$. The most sensitive
determination, by far, is based upon 
$\pi N$ elastic analysis, with fixed-t
and forward dispersion relation constraints. This yields uncertainties 
of $(\pm 0.01)$ (mapping error) + $(0.03)$ (extraction error). Next in
sensitivity are the combined $NN$ analyses to 500 MeV, where the 
uncertainty is estimated to be $(\pm 0.13)$ (for SP40) and 
$(\pm 0.09)$ (for SG40). Least in sensitivity was pion-photoproduction,
where a fit to 500 MeV gave an uncertainty of $(\pm 0.13)$ and where
an average taken over various maximum energies gave only a slightly
different sensitivity ($\pm 0.14)$. 

Clearly the $\pi N$ elastic reaction shows a far greater sensitivity and,
in our opinion, provides the least model-dependent determination.

\acknowledgments
This work was supported in part by the U.~S.~Department of Energy Grants
DE--FG02--99ER41110, DE--FG02--97ER41038 and DE--FG02--95ER40901.


\newpage
{\Large\bf Figure captions}\\
\newcounter{fig}
\begin{list}{Figure \arabic{fig}.}
{\usecounter{fig}\setlength{\rightmargin}{\leftmargin}}
\item
{ Extracted values of $g^2 /4 \pi$ using different upper limits
for the lab photon energy (E$_{\gamma}$). See text. }

\end{list}

\newpage
\mediumtext
\vfill
\eject
Table~I. 
Fit SP40 from 0-400 MeV (57 parameters, J$_{\rm max}$=6) including
Uppsala data. Results are for $pp$ and $np$ data, Uppsala measurements
at 96 MeV (U96) and 162 MeV (U162). The mapping in
$\chi^2$ is calculated with and without (columns PPx and NPx) the 
$\chi^2$ contribution from Uppsala data. The fit incorporates 3421
$pp$ data and 3856 $np$ data (U96: 53 data, U162: 54 data).
\vskip .5cm
\centerline{
\begin{tabular}{cccccccc}
\hline
\noalign{\vskip 10pt}
$g^2/ 4\pi$ & $\chi^2_{pp}$ & $\chi^2_{np}$ & $\chi^2_{\rm u96}$ & & 
 $\chi^2_{\rm u162}$ & $\chi^2_{PPx}$ & $\chi^2_{NPx}$ \\ 
\noalign{\vskip 10pt}
\hline
\noalign{\vskip 10pt}
13.50 & 4377  & 5440 & 92 & & 255 & 4371 & 5000 \\
\noalign{\vskip 6pt}
13.75 & 4375  & 5415 & 89 & & 252 & 4368 & 4980 \\
\noalign{\vskip 6pt}
14.00 & 4386  & 5399 & 87 & & 249 & 4379 & 4968 \\
\noalign{\vskip 6pt}
14.25 & 4411  & 5291 & 85 & & 246 & 4404 & 4963 \\
\noalign{\vskip 6pt}
14.50 & 4450  & 5391 & 83 & & 243 & 4443 & 4966 \\
\noalign{\vskip 6pt}
\hline
\noalign{\vskip 10pt}
$g^2 /4 \pi_{\rm min}$ & 13.67(10) & 14.37(13) & --- & & --- 
 & 13.68(10) & 14.28(13) \\
\noalign{\vskip 10pt}
\hline
\end{tabular} }
\vfill
\eject
Table~II.
Fit SG40 from 0-400 MeV (49 parameters, J$_{\rm max}$=4) including
Uppsala data. Notation as in Table~I. 
\vskip .5cm
\centerline{
\begin{tabular}{cccccccc}
\hline
\noalign{\vskip 10pt}
$g^2/ 4\pi$ & $\chi^2_{pp}$ & $\chi^2_{np}$ & $\chi^2_{\rm u96}$ & &
 $\chi^2_{\rm u162}$ & $\chi^2_{PPx}$ & $\chi^2_{NPx}$ \\
\noalign{\vskip 10pt}
\hline
\noalign{\vskip 10pt}
13.50 & 4510  & 5475 & 89 & & 251 & 4512 & 5038 \\
\noalign{\vskip 6pt}
13.75 & 4513  & 5470 & 85 & & 249 & 4515 & 5037 \\
\noalign{\vskip 6pt}
14.00 & 4533  & 5482 & 82 & & 246 & 4535 & 5052 \\
\noalign{\vskip 6pt}
14.25 & 4569  & 5512 & 80 & & 244 & 4572 & 5085 \\
\noalign{\vskip 6pt}
14.50 & 4622  & 5559 & 78 & & 242 & 4626 & 5134 \\
\noalign{\vskip 6pt}
\hline
\noalign{\vskip 10pt}
$g^2 /4 \pi_{\rm min}$ & 13.56(9) & 13.69(9) & --- & & ---
 & 13.58(9) & 13.64(9) \\
\noalign{\vskip 10pt}
\hline
\end{tabular} }
\vfill
\eject
Table~III.
Fit SX40 from 0-400 MeV (48 parameters, J$_{\rm max}$=4) excluding
$np$ charge-exchange cross section data. Fit includes 
3421 $pp$ data and 1683 $np$ data (see text).
\vskip .5cm
\centerline{
\begin{tabular}{ccc}
\hline
\noalign{\vskip 10pt}
$g^2/ 4\pi$ & $\chi^2_{pp}$ & $\chi^2_{np}$ \\
\noalign{\vskip 10pt}
\hline
\noalign{\vskip 10pt}
13.50 & 4507  & 2216  \\
\noalign{\vskip 6pt}
13.75 & 4511  & 2200  \\
\noalign{\vskip 6pt}
14.00 & 4530  & 2192  \\
\noalign{\vskip 6pt}
14.25 & 4566  & 2188  \\
\noalign{\vskip 6pt}
14.50 & 4618  & 2190  \\
\noalign{\vskip 6pt}
\hline
\noalign{\vskip 10pt}
$g^2 /4 \pi_{\rm min}$ & 13.57(9) & 14.28(15) \\
\noalign{\vskip 10pt}
\hline
\end{tabular} }
\vfill
\eject
Table~IV.
Fit to pion-photoproduction data (threshold to 500 MeV)
for different values of $g^2/ 4\pi$ (see text). Fit includes
3441 $\pi^0 p$ data, 2416 $\pi^+ n$ data, and 
974 $\pi^- p$ data.
\vskip .5cm
\centerline{
\begin{tabular}{ccccc}
\hline
\noalign{\vskip 10pt}
$g^2 /4\pi$ & $\chi^2_{\rm all}$ & $\chi^2_{\pi^0 p}$ & 
$\chi^2_{\pi^+ n}$ & $\chi^2_{\pi^- p}$ \\
\noalign{\vskip 10pt}
\hline
\noalign{\vskip 10pt}
13.50 & 13084 & 7457 & 3723 & 1865 \\
\noalign{\vskip 6pt}
13.75 & 13074 & 7465 & 3727 & 1843 \\
\noalign{\vskip 6pt}
14.00 & 13071 & 7473 & 3735 & 1824 \\
\noalign{\vskip 6pt}
14.25 & 13075 & 7482 & 3745 & 1808 \\
\noalign{\vskip 6pt}
14.50 & 13088 & 7493 & 3760 & 1795 \\
\noalign{\vskip 6pt}
\hline
\noalign{\vskip 10pt}
$g^2 /4\pi_{\rm min}$ & 13.97(13) & --- & --- & --- \\
\noalign{\vskip 10pt}
\hline
\end{tabular} }
\vfill
\eject

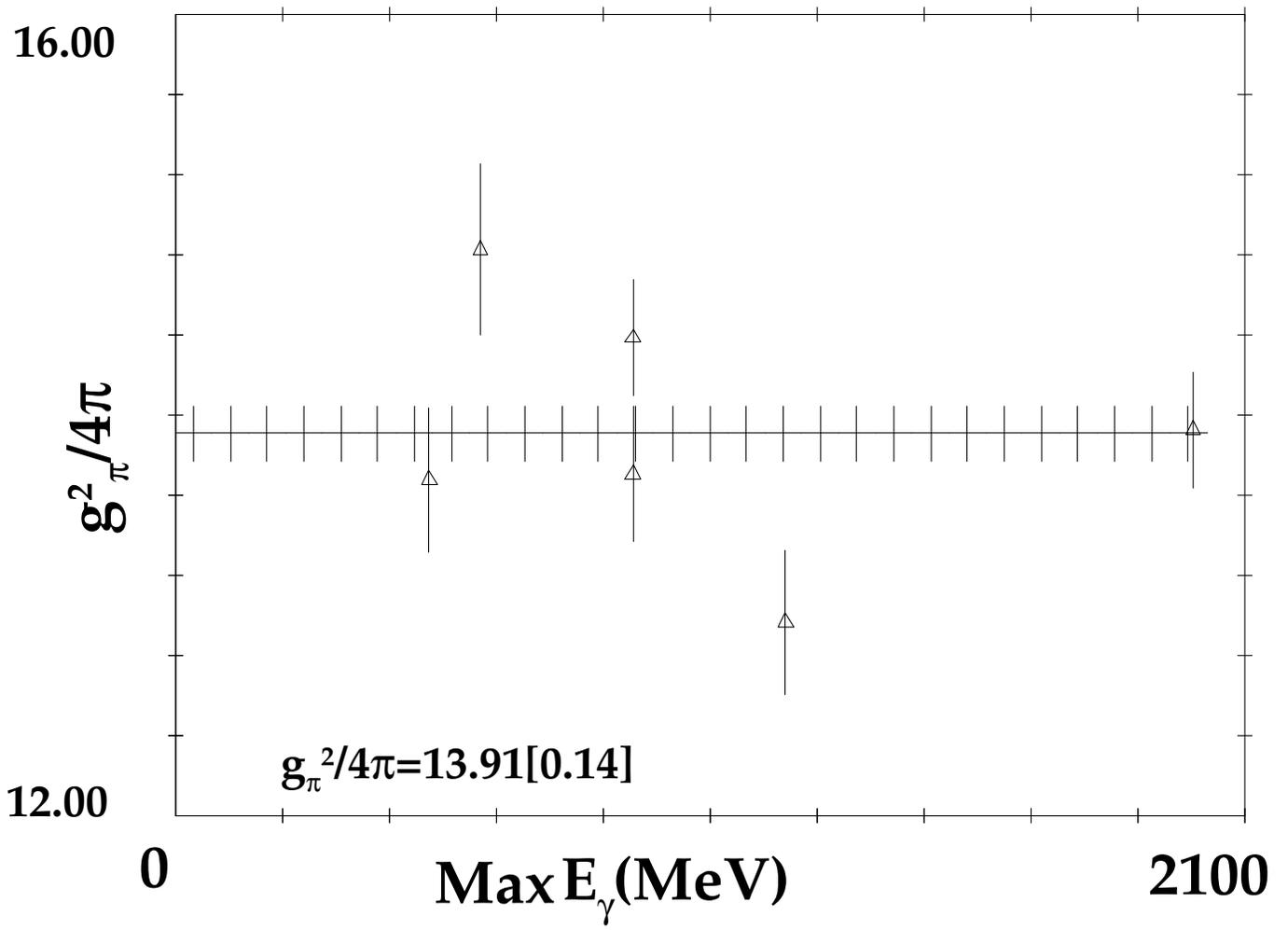
\end{document}